\documentclass{pasj00}
\draft

\begin{document}
\SetRunningHead{T. Hioki et al.}{Optical/Near-IR Observations of FS Tau Binary}
\Received{2000/12/31}
\Accepted{2001/01/01}

\title{High-Resolution Optical and Near-Infrared Images of 
the FS Tauri Circumbinary Disk\footnotemark[\ast]}

\footnotetext[$\ast$]{Based on data collected at the Subaru Telescope, 
which is operated by the National Astronomical Observatory of Japan.}
%
\author{Tomonori \textsc{Hioki}\altaffilmark{1},
        Yoichi \textsc{Itoh}\altaffilmark{1}, 
        Yumiko \textsc{Oasa}\altaffilmark{2},
        Misato \textsc{Fukagawa}\altaffilmark{3},
        and
	Masahiko \textsc{Hayashi}\altaffilmark{4}
}
\altaffiltext{1}
{Graduate School of Science and Technology, Kobe University, 
1-1 Rokkodai, Nada, Kobe 657-8501}
\email {yitoh@kobe-u.ac.jp}
\altaffiltext{2}
{Faculty of Education, Saitama University, 
255 Shimookubo, Sakura, Saitama, Saitama 338-0825}
\altaffiltext{3}
{Graduate School of Science, Osaka University, 
1-1 Machikaneyama-cho, Toyonaka, Osaka 560-0043}
\altaffiltext{4}
{Graduate School of Science, The University of Tokyo, 
7-3-1 Hongo, Bunkyo, Tokyo 113-0033}

\KeyWords{star: individual (FS Taurus) --- star: pre-main sequence --- techniques: high angular resolution}

\maketitle

\begin{abstract}
We present an $H$-band image of FS Tauri, 
a 0\farcs2-separated classical T Tauri binary system, 
taken with the Coronagraphic Imager with Adaptive Optics (CIAO) 
on the Subaru Telescope. 
This new image, combined with Hubble Space Telescope / Advanced Camera
for Surveys ($HST$/ACS) F606W-band polarimetric images, 
shows that the binary has complicated circumbinary features, 
including a circumbinary disk, western and eastern arm-like structures, 
and two cavities at the northeast and southwest.
The circumbinary disk is 630 AU in radius and the southeast side of 
the disk is bright in the $H$-band. 
The brightness ratio (southeast/northwest) is $1.6\pm0.4$. 
A single Rayleigh-like scattering model indicates that the 
disk is inclined by 30\degree~ to 40\degree~ and that
the southeast side corresponds to the near side along our line of sight. 
The $H$-band surface brightness of the southeast side 
decreases as $r^{-1.9\pm0.1}$ from
15.2 mag arcsec$^{-2}$ to 16.8 mag arcsec$^{-2}$.
The outer portion of the disk is possibly more flared than its inner portion. 
The weak centro-symmetric polarization pattern and redder F606W$-H$ color 
($4.2\pm0.2$ mag) of the southeast side are probably caused by 
multiple scattering events from the dust grains associated with the binary. 
The F606W-band image shows the bright northwest side of the disk in 
contrast with the $H$-band image. 
The F606W$-H$ color of the northwest side is between 1.7 mag and 3.0 mag. 
We consider that Haro 6-5 B (FS Tauri B), 20\arcsec~ away,
produces the neutral scattered light from the northwest side. 
This idea is supported by the polarization pattern of the northwest 
side, which is centro-symmetric with respect to Haro 6-5 B. 
The arms appear to encompass the western and eastern cavities, 
suggesting that the arms + cavity systems are created by a bipolar 
outflow from the binary. 
However, the direction of this outflow is inconsistent with that of 
outflows inferred from the circumbinary disk model. 
These differences may arise from misalignment between the circumbinary 
disk and the circumstellar disks. 
Another mechanism forming the arms + cavity systems is considered 
to be the inhomogeneous density distribution of materials 
in the circumbinary disk.
\end{abstract}

\section{Introduction}
Many studies have showed that the majority of young stars ($\leq 1$ Myr) are
associated with various circumstellar structures, 
such as envelopes, protoplanetary disks, and/or bipolar outflows. 
The disks are composed of gas and dust, the possible birthplaces of planets.
The bipolar outflow creates a cavity in the envelope. 
The cavity allows optical and near-infrared (NIR) radiation
to escape and illuminate an extended reflection nebula. 
These circumstellar features have been directly revealed using 
ground-based telescopes with adaptive optics (AO) and 
the $Hubble~Space~Telescope$ ($HST$). 

A significant number of young stars appear in binary systems. 
NIR imaging surveys show multiplicity rates of $29-37$\% for 
Class I proto-stars in the $\rho$ Ophiuchi molecular cloud 
(separation $<$ 2000 AU; Haisch et al. 2002, 2004; Duch\^ene et al. 2004). 
This rate is quite high, despite the relatively poor sensitivity 
at the small projected separations ($< 110$ AU). 
Some mechanisms are proposed for the formation of protobinaries,
in which the core fragmentation process is important (e.g., Boss 2002). 
In the standard core fragmentation senario, 
a rotational and magnetic cloud core fragments into multiple dense objects 
and then a proto-binary is formed.
The semi-major axis of the binary is determined by the initial core properties, 
especially by the magnetic and rotational energies and the angular momentum 
(Machida et al. 2008). 

In the more evolved stage (Class II or classical T Tauri stars), 
binary systems have two kinds of protoplanetary disks: 
circumstellar disks associated with each star and ring-shaped 
circumbinary disks around the binary systems (Artymowicz \& Lubow 1994, 1996).
The sizes of the circumbinary disks and circumstellar disks depend 
on the binary system's semimajor axis, eccentricity, and mass ratio 
(Artymowicz \& Lubow 1996; Armitage et al. 1999). 
Several observations have detected scattered light from the circumbinary 
disks using optical and NIR imaging/polarimetric methods, 
for instance, around the GG Tau (Itoh et al. 2002; McCabe et al. 2002; 
Krist et al. 2005; Silber et al. 2000) and UY Aur (Close et al. 1998; 
Hioki et al. 2007; Potter et al. 2000) binary systems. 
Bipolar outflows also appear in several T Tauri binary systems 
(e.g., Hirth et al. 1997). 
However, observations have focused mainly on single T Tauri stars, 
although more than half of all T Tauri stars are binaries, according to
observational results (separation $<$ 1800 AU;
Ghez et al. 1993; Leinert et al. 1993) 
and theoretical predictions 
(e.g., Nakamura \& Li 2003; Machida et al. 2008). 
A limited number of studies have examined the structures around 
T Tauri binary systems. 

FS Tauri A (hereafter FS Tau or FS Tau binary) is a classical T Tauri 
binary system in the Taurus molecular cloud ($d \sim 140$ pc; Elias 1978). 
The primary and secondary stars have spectral types of M0 and M3.5, 
respectively, with a projected separation, $a_{\mathrm{sep}}$, 
of 0.242\arcsec~ (34 AU; Hartigan \& Kenyon 2003). 
The dynamical mass of the binary is calculated to be $0.78\pm0.25$ \MO~ 
(Tamazian et al. 2002). 
Based on the shape of the spectral energy distribution (SED), 
the binary is classified between Class I and II (Andrews \& Williams 2005). 

Gledhill \& Scarrott (1989) suggested a circumbinary disk with a diameter 
of $\sim$ 5\arcsec~ (700 AU). 
The disk shows a linear polarization of $\sim$ 10 \% in the $R$-band, 
which is higher than that of its associated molecular cloud 
($\sim$ 2 \%; Vrba et al. 1985). 
However, Krist et al. (1998) could not confirm a disk beyond 60 AU from 
the binary using $HST$ / Wide Field Planetary Camera 2 (WFPC2) observations 
at the $V$-, $R$-, and $I$-band wavelengths, 
although they did not subtract the point-spread functions (PSFs) of the binary. 
The sub-millimeter and millimeter continuum fluxes of the binary are 
extremely low in comparison with those of other T Tauri stars 
(Andrews \& Williams 2005; Beckwith et al. 1990; Dutrey et al. 1996). 
The gas and dust mass around the binary is calculated to be 0.002 \MO~. 

FS Tau is actually a hierarchical triple system.
The third component, called as FS Tau B or Haro 6-5 B, is located about 
20\arcsec~ (2800 AU) west of the FS Tau binary.
However, high spatial resolution imaging observations revealed no directly visible star in the Haro 6-5 B system (Krist et al. 1998).
Haro 6-5 B itself is obscured by its circumstellar disk, which
is detected in millimeter wavelengths (Dutrey et al. 1996).
The companionship of Haro 6-5 B to FS Tau A is not proven
by common proper motion, spectroscopy, nor photometry.
Instead, the companionship is indicated by polarimetry.
The FS Tau A binary is surrounded by a reflection nebula up to 
1400 AU in radius, 
which shows a centro-symmetric polarization pattern centered on the binary, 
particularly on the south side (Gledhill \& Scarrott 1989). 
Krist et al. (1998) suggested that the north side is illuminated not only 
by the FS Tau binary but also by Haro 6-5 B.
H$_{\alpha}$ imaging observations indicated a cavity wall in the north side, 
excited by the blue-shifted outflow from Haro 6-5 B 
(Eisl$\ddot{\mathrm{o}}$ffel \& Mundt 1998). 
The reflection nebula and the cavity wall appear to be truncated 
by the straight edge of a dark cloud $\sim 700$ AU north of the binary 
(Fig. \ref{HST}; see also Fig. 3 of Krist et al. 1998). 

The FS Tau binary has not been known to drive outflows. 
Optical emission lines, such as [S II] and H$_{\alpha}$, showed no 
evidence of high-velocity jets from the binary (Woitas et al. 2002). 
Although $^{13}$CO emission was observed along the southeast direction 
from the binary, it is unclear whether this emission arises from a diffuse 
wind (Dutrey et al. 1996). 

In this paper, we present a subarcsecond NIR image of the FS Tau binary 
taken with the Coronagraphic Imager with Adaptive Optics (CIAO) 
on the Subaru Telescope. 
Combined with the $HST$ optical imaging and polarimetric methods, 
we discovered complex features associated with the binary. 
Our goal was to dissect the circumbinary disk, whose southeast side 
is bright in NIR wavelengths, in contrast with the northwest side,
which is bright in the optical. 
We interpret the difference in brightness as being caused by the
forward scattering of dust 
on the southeast side and a contribution to the radiation from Haro 6-5 B 
on the northwest side. 
Observations and data reduction are described in $\S 2$. 
Our results and some 
discussion are presented in $\S 3$ and $\S 4$, respectively. 
Finally, our conclusions are presented in $\S 5$. 

\section{Observations and Data Reduction} 

\subsection{CIAO} 

Near-infrared ($H$-band; 1.6 \micron~) coronagraphic observations of 
the FS Tau binary were carried out on 2007 December 17 with the 
Subaru/CIAO (Tamura et al. 2000) system.
A 0\farcs8 mask blocked the light from both the primary and secondary stars 
($a_{\mathrm{sep}} \sim$ 0\farcs24). 
This mask had $\sim$ 2\% transmission (Murakawa et al. 2003). 
A pupil Lyot stop reduced the diffracted light. 
The field of view was 21\farcs83 $\times$ 21\farcs83, with a pixel scale of 
0\farcs02132$\pm$0\farcs00003 pixel$^{-1}$ and a position angle (P.A.) 
on the detector of 1\fdg58$\pm$0\fdg07. 
The pixel scale and orientation were measured with observations of 
the Trapezium cluster (Simon et al. 1999). 
One may doubt whether the astrometrical calibration is still valid,
since the data of Simon et al. (1999) were taken in 1996.
The pixel scale and the P.A. derived here are consistent with those derived
by Itoh et al. (2005), whose data were taken in 2003.
Thus we believe that these parameters are correctly derived.
During the observations, the natural seeing size varied
between 0\farcs9 and 1\farcs4 in the $H$-band. 
The average full width at half-maximum (FWHM) of the PSFs of our observations 
was 0\farcs4, using AO. 
We obtained 36 object frames in the $H$-band. 
The integration time was 10 s $\times$ 3 co-adds for each object frame. 
After obtaining the first 24 frames, we performed dithering to 
remove hot and bad pixels. 
We did not obtain science -- sky -- science jittering
as it is the standard for infrared observations of extended sources.
This is due to very low observing efficiency.
For the coronagraphic observations with the AO system,
the setup procedures are very complicated,
as about 10 minutes was required for every pointing.
After the FS Tau observations, 
we also obtained 25 frames of a PSF reference star, SAO 76648, 
in the $H$-band using the 0\farcs8 mask. 
It is located about 140\arcmin~ east to FS Tau.
The $H$-band magnitude of the PSF reference star was only 0.5 mag 
brighter than that of FS Tau. 
On the other hand, the $R$-band magnitude for the wavefront sensing
is 1.5 mag brighter than that of FS Tau.
We adjusted it to match FS Tau using neutral density (ND) 
filters in the AO.
This was necessary to make the PSF of the reference star similar to that of 
FS Tau. 
The integration time was 10 s $\times$ 3 co-adds for each reference star frame.
FS 123 was observed as a photometric calibrator (Hawarden et al. 2001). 
Twilight-flats and dark frames were taken at the end of the night. 

The object frames were calibrated using the Image Reduction and 
Analysis Facility (IRAF). 
First, a dark frame was subtracted from the object frames. 
Then the frames were divided by the normalized twilight-flat to 
remove pixel-to-pixel variation in sensitivity. 
Hot and bad pixels were removed from the divided frames using the 
\texttt{COSMICRAYS} task. 
Then, an averaged count in the sky region was subtracted. 
We judged that neither halo of the PSF nor circumstellar structures
extended at 6\arcsec~ away from the central star.
The count was measured along a concentric circle (radius of $\sim$ 6\arcsec~
with a width of $\sim$ 1\arcsec~) on the FS Tau binary, 
avoiding the diffraction artifacts of the spider. 
The same reduction procedures were applied to the reference star frames. 

We subtracted the reference star frames from the object frames 
to detect faint structures buried in the halo of the central binary. 
The peak positions of the FS Tau binary and reference star were measured 
with the \texttt{IMEXAMINE} task on all frames. 
For FS Tau, the peak positions were able to be measured for 21 frames.
For the rest frames, we did not identify the core of the PSF.
The secondary star (FS Tau Ab) was not resolved in the object frames 
because the averaged FWHM was about 0\farcs4. 
We assumed that the measured positions of the PSF peak corresponded
to the positions of the primary. 
The object frames were shifted so that the positions of the primary 
were centered on the image. 
The reference star frames were duplicated after dark-subtraction, 
flat-fielding, hot and bad pixel rejection, and sky-subtraction. 
Then each reference star frame was shifted to adjust the peak positions 
of the PSFs between the reference star and each component of the binary. 
We calculated the positions of the secondary from the separation and P.A. 
in the $HST$/ACS image (see $\S 3.2$). 
As the binary has only a separation of 0\farcs23 (32 AU)
and its total mass is 0.78 \MO~ this yields an orbital period
of about 210 years.
Hence, the position angle of the binary could have changed
up to 5\degree~ within the epoch difference of 3 years between
the ACS and CIAO observations.
It produces 1 pixel uncertainty (0\farcs23 $\times \tan$ 5\degree~) 
in the position of the secondary star.
The intensities of the reference star frame were scaled to adjust 
the flux ratio of the binary. 
Because the $H$-band flux ratio of the binary is unknown, 
we presumed that it was the same as the $K$-band flux ratio 
($7.0\pm0.1$; White \& Ghez 2001).
One may consider that the $H$-band flux ratio can be calculated from
the spectral types of Aa and Ab and the $K$-band flux ratio.
However, since FS Tau has near-infrared excess ($J=10.705\pm0.027$ mag,
$H=9.244\pm0.026$ mag, $K=8.178\pm0.017$ mag from 2MASS) and we cannot
identify whether the excess arises from FS Tau Aa and/or Ab,
we did not determine the flux ratio of Aa and Ab in the $H$-band.
We are convinced from the final PSF-subtracted image that the assumed flux
ratio is appropriate.
Then the 12 reference star frames, except the frames with the asymmetric PSFs,
were combined. 
The combined reference star frames were renormalized to match the 
intensities between the reference star frames and the object frames 
in the 0\farcs5$\times$0\farcs5 region immediately outside the mask 
southwest of the binary, where the effect of the spider was negligible. 
The reference star frames were subtracted from each object frame. 
Finally, the 21 PSF-subtracted object frames, except the frames.
We checked the photometric accuracy under the unstable seeing condition.
Any object except for FS Tau itself did not appear in the object
frames.
FS Tau in the object frames is not appropriate for evaluating 
the photometric accuracy, because the core of the PSF was under the
mask.
Instead, we evaluated it with the $R$-band flux of FS Tau, which was
monitored during the integrations by the wavefront sensor in the AO system.
The measured fluxes were stable with its standard deviation
of 0.04 mag.
We consider that our observations have photometric accuracy of 0.04 mag.

\subsection{HST Polarimetric Data} 

Polarimetric observations were carried out on 2004 August 21 
with the $HST$ Advanced Camera for Surveys (ACS). 
These observations were proposed by D. Hines ($HST$ program GO 10178). 
The central wavelength and the width of the F606W filter (broad $V$-band) 
are 5907 \AA~ and 2342 \AA~, respectively. 
The field of view was 202\arcsec~$\times$202\arcsec~, 
with a pixel scale of 0\farcs050$\pm$0\farcs005 pixel$^{-1}$
(The ACS manual). 
The position angles of the images were recorded in the FITS keyword
of the image, ORIENTAT.
The uncertainty in the position angle was 0\fdg01 -- 0\fdg03 in the
worst cases (Koekemoer et al. 2007).
The exposure time of each polarized image (0\degree~, 60\degree~, and 120\degree~)
was 536 s. 
The averaged FWHM of the observations was 0\farcs09. 

We downloaded the polarimetric data with the basic calibration from
the HST MAST web site.
Each frame was shifted so that the positions of the unsaturated FS Tau 
primary were centered on the image. 
The Stokes parameters, $I$, $Q$, and $U$ were computed using
\begin{eqnarray}
I &=& \frac{2}{3}(F_{0}+F_{60}+F_{120})\\
Q &=& \frac{2}{3}(2F_{0}-F_{60}-F_{120})\\
U &=& \frac{2}{\sqrt{3}}(F_{60}-F_{120}),
\end{eqnarray}
where $F_{0}$, $F_{60}$, and $F_{120}$ are the counts of 
each polarized image (Biretta et al. 2004). 
The degree of linear polarization $P$ and the polarization angle 
$\theta$ were computed from
\begin{eqnarray}
P &=& \frac{\sqrt{Q^2+U^2}}{I}\\
\theta &=& \frac{1}{2}\arctan({\frac{U}{Q}}).
\end{eqnarray}
The uncertainty in the degree of polarization was $\sim$ 0.3\%, 
and the uncertainty in polarization angle was $<$ 2\fdg0.

We subtracted the PSFs of the primary and secondary stars in the Stokes $I$ 
image from those created by the Tiny Tim program (model PSFs; Krist 2004) 
to detect circumbinary structures using the same data reduction 
as CIAO.
The intensities of the model PSFs were scaled to adjust the $V$-band 
flux ratio of the binary (20.8; Krist et al. 1998). 
We combined the model PSF frames, and then the combined frame was 
renormalized to match the intensities between the model PSF frames and 
the object frames of the 0\farcs5$\times$0\farcs5 southwest region of 
the binary, immediately outside the "artifact region" (see \S \ref{Vimage}). 

\section{Results}

\subsection{$H$-band Coronagraphic Image}

\begin{figure}
\begin{center}
\FigureFile(80mm,80mm){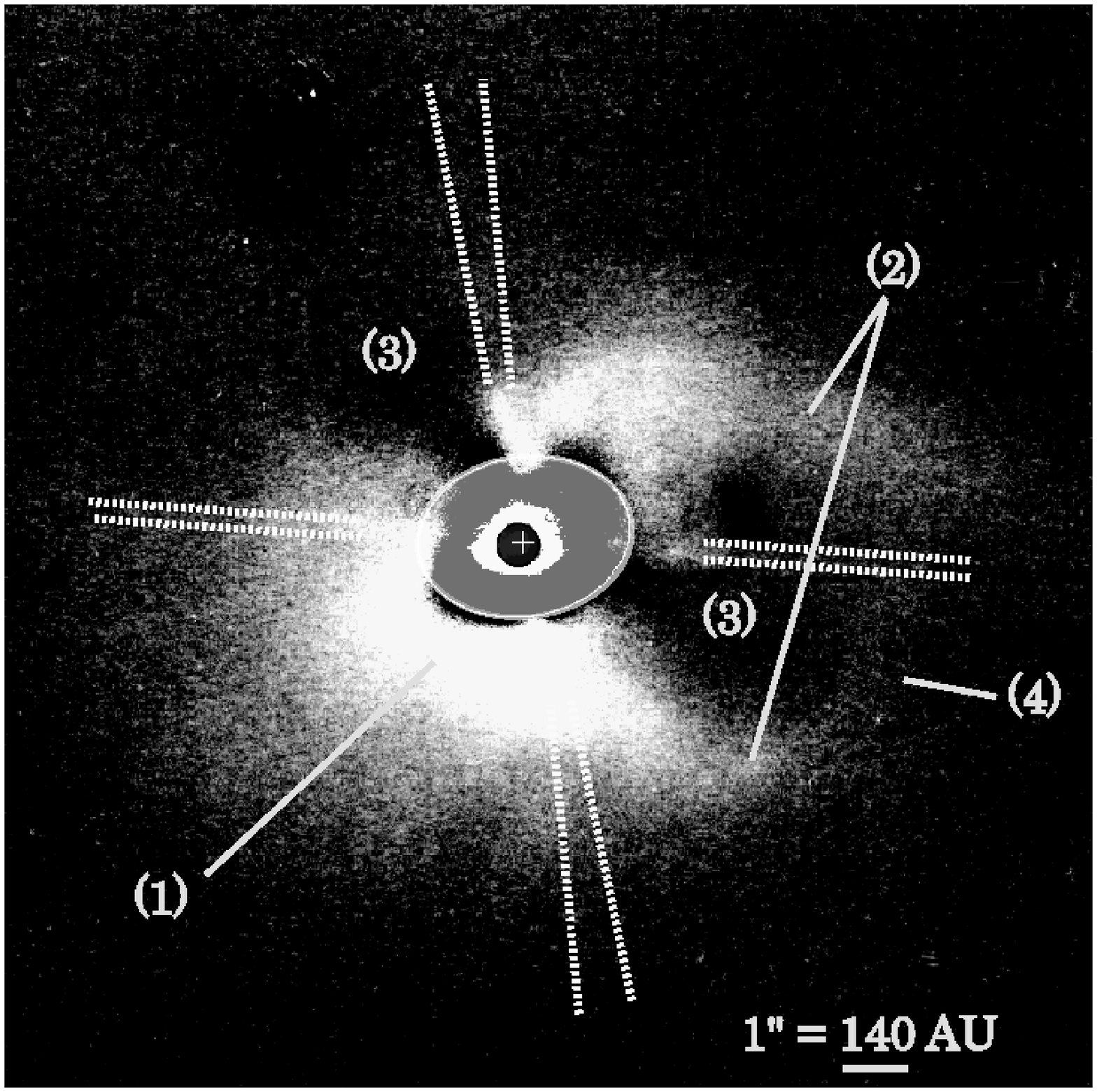}
\FigureFile(80mm,80mm){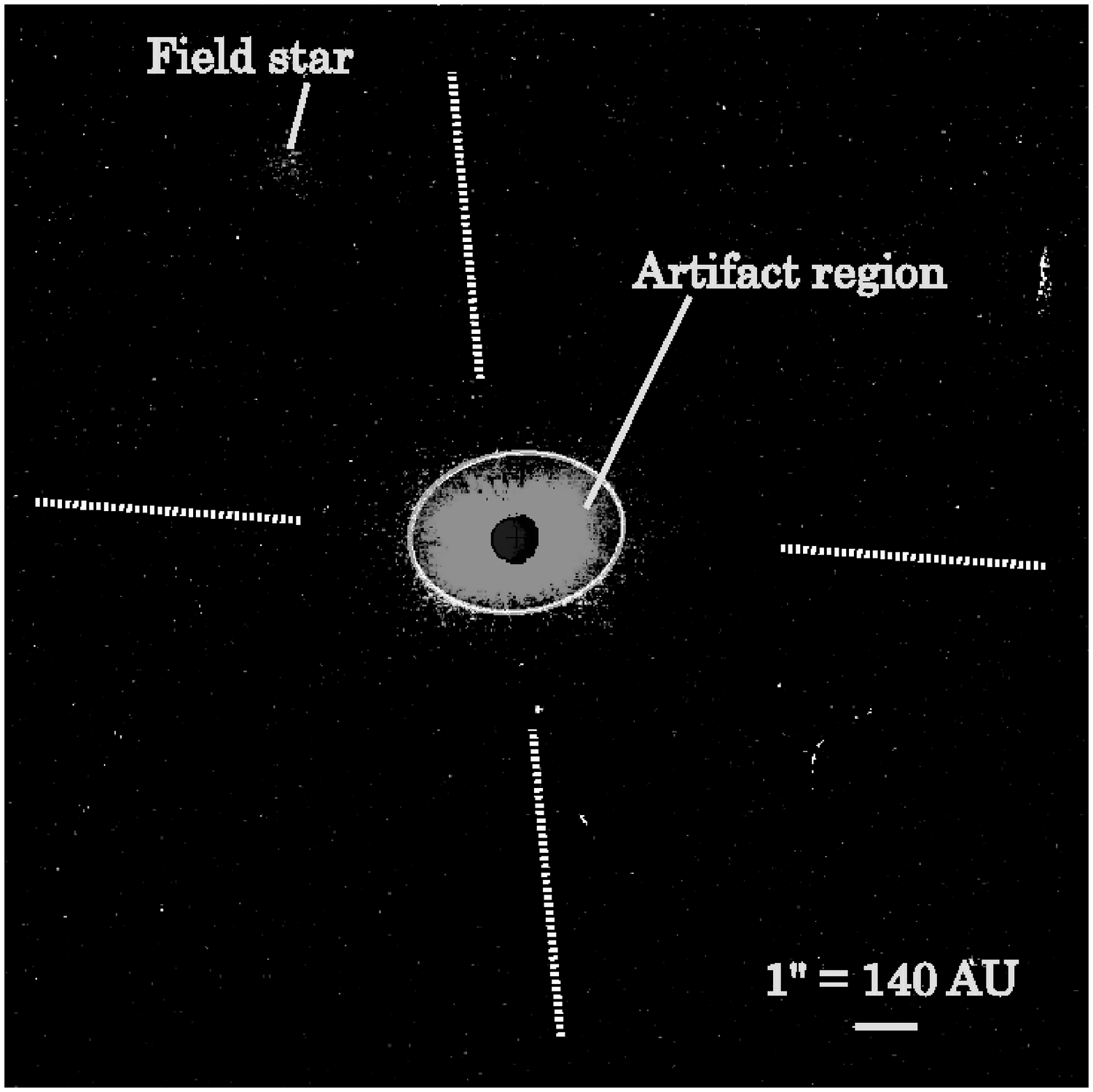}
\caption{The $H$-band coronagraphic image of the FS Tau binary system ($left$). 
The PSFs of the central binary were subtracted. 
"1" indicates the $H$-band bright portion of the circumbinary disk; 
"2" indicates the arm-like structures; 
"3" indicates the cavities; and 
"4" indicates the arc-like structure. 
The position of the FS Tau primary is indicated by the cross symbol in 
the occulting mask with a diameter of 0\farcs8 (112 AU; a circular symbol). 
The binary is not resolved due to bad seeing conditions. 
The pseudo-binary was created using the PSF reference star SAO 76648 
($right$). 
The PSF-subtracted image shows some artifacts within 1\farcs3 - 1\farcs7
from the central star, represented as hatched ellipses in both images. 
Beyond this region, the diffraction pattern of the spider is the only artifact,
indicated by dotted lines in the both images. 
The field of view (FOV) is 17\farcs8$\times$17\farcs8. 
North is up and east is to the left.} 
\label{CIAO}
\end{center}
\end{figure}

The $H$-band coronagraphic image is shown in Fig. \ref{CIAO}. 
Despite bad seeing conditions an image with 0\farcs4 FWHM could be obtained.
The FS Tau binary exhibited four remarkable structures: 
\begin{enumerate}
\item 
A bright portion extending
630 AU southeast from the binary ("1" in Fig. \ref{CIAO}). 
\item 
Two arm-like structures on the western side ("2" in Fig. \ref{CIAO}). 
\item 
Two cavities at the northeast and southwest at P.A. $\sim$20\degree~ and 
$\sim$250\degree~, respectively ("3" in Fig. \ref{CIAO}). 
\item 
A faint ($>$17.3 mag arcsec$^{-2}$) arc-like structure $\sim$900 AU 
west of the binary ("4" in Fig. \ref{CIAO}). 
\end{enumerate}

We measured the surface brightness of the extended features with 
0\farcs2 $\times$ 0\farcs2 square apertures.
The southeast bright structure has 15.2$\pm$0.3 mag arcsec$^{-2}$ at 220 AU
from the central star and 16.8$\pm$0.2 mag arcsec$^{-2}$ at 530 AU.
If the PSF subtraction of the reference star were too strong or too weak, 
the residual would be seen as a circle or streak. 
We confirmed that the detected structures are real by making a pseudo-binary 
frame from the first half of the reference star frames. 
The projected separation and flux ratio of the pseudobinary were
the same as those of the FS Tau binary ($\S 2.1$). 
We subtracted the PSFs using the same procedure as for FS Tau. 
In the processed image, we identified subtraction residuals
within $1\farcs3 - 1\farcs7$ 
from the central star (Fig. \ref{CIAO}). 
We called this region the "artifact region."
Beyond this region, the diffraction pattern of the spider was the only artifact.
We determined the limiting magnitude of the observations by measuring
standard deviation of the sky regions which was set far from
the central star and the nebulosity.
We found that the limiting magnitude with S/N=3 was 17.3 mag arcsec$^{-2}$.

\subsection{F606W-band Stokes $I$ Image}
\label{Vimage}

\begin{figure}
\begin{center}
\FigureFile(85.3mm,85.3mm){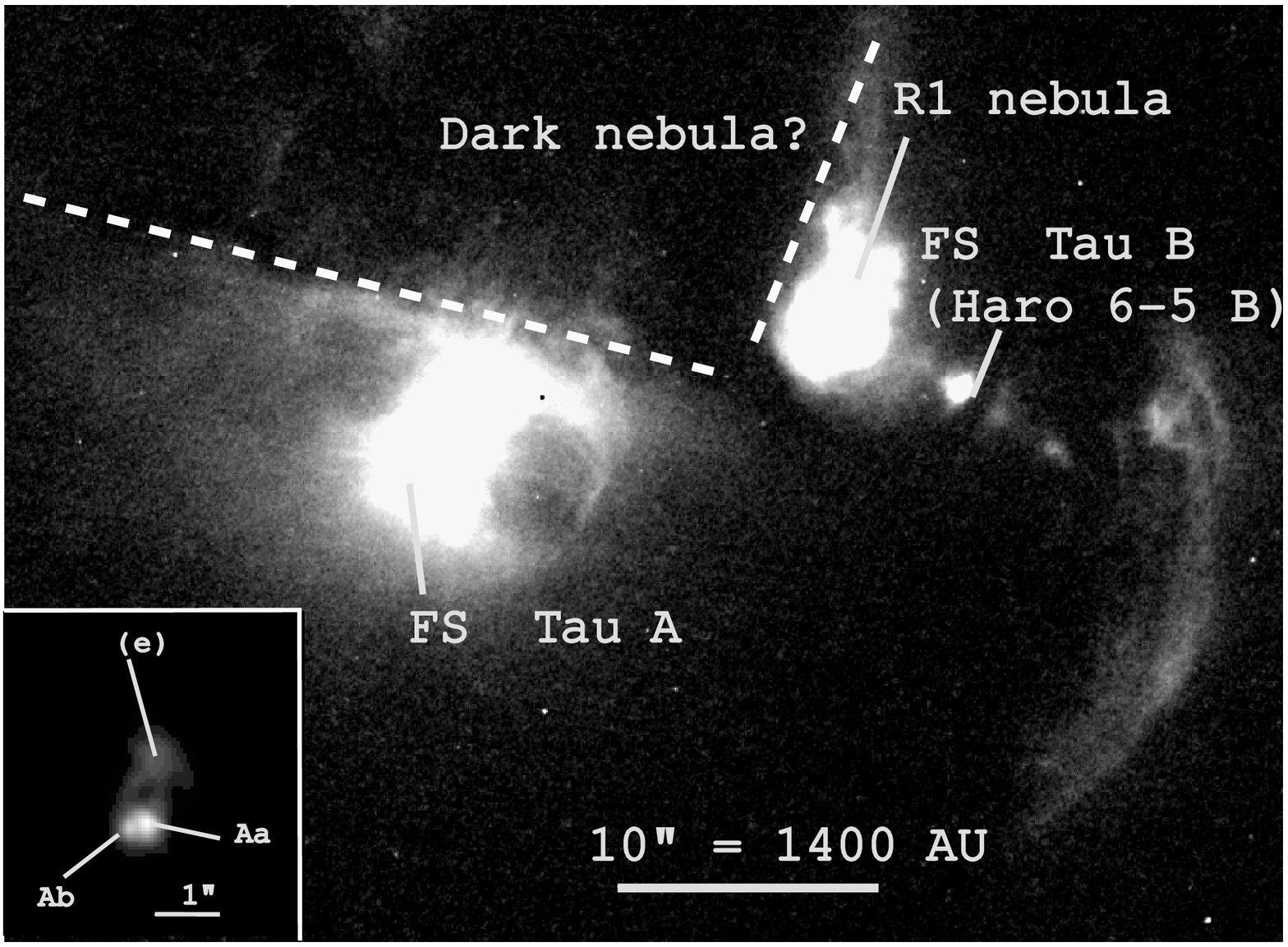}
\FigureFile(76mm,76mm){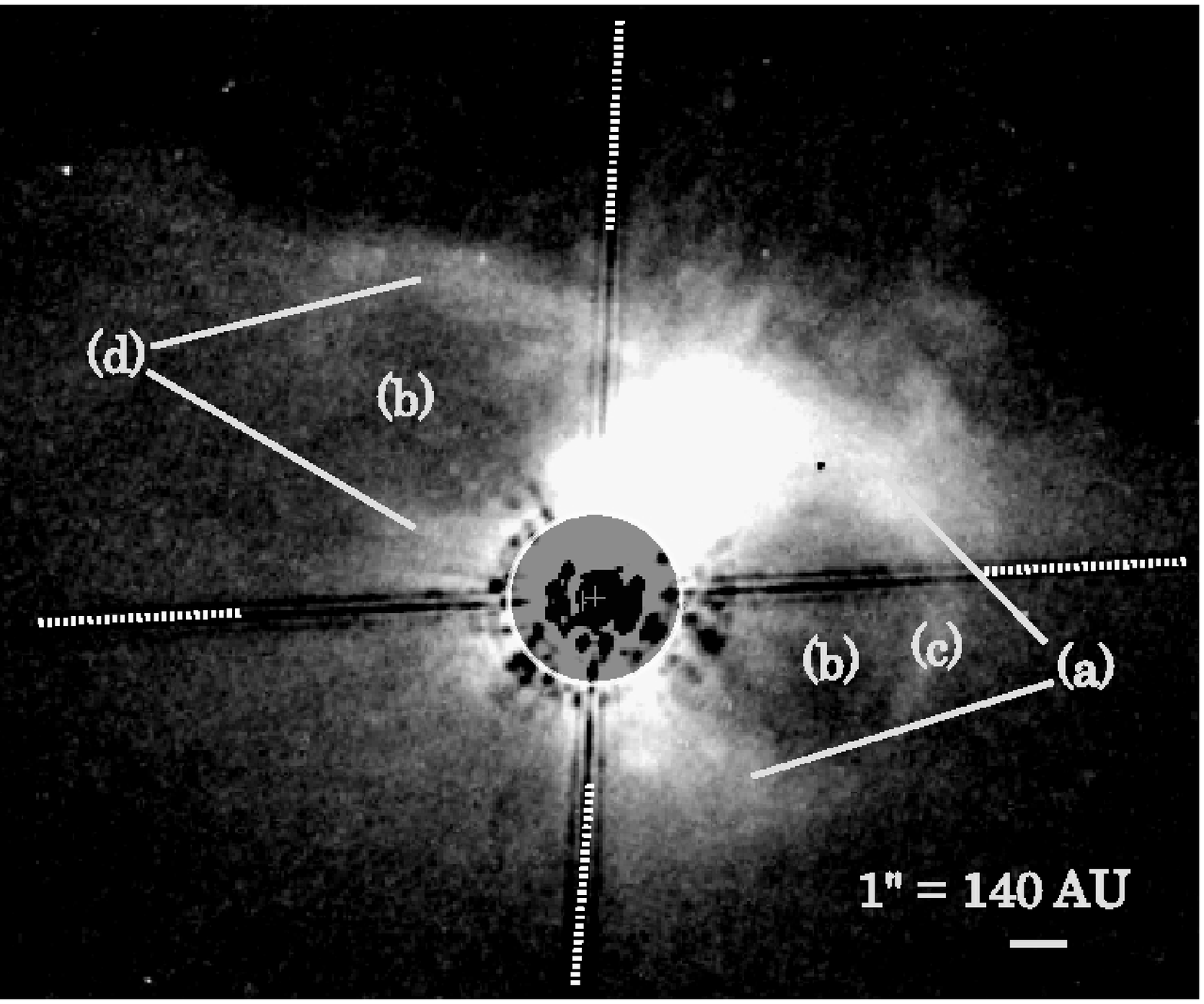}
\caption{The F606W-band Stokes $I$ images of the FS Tau/Haro 6-5 B 
field before subtracting the PSFs of the central binary ($left$) 
and a closeup of the binary after subtracting the PSFs ($right$). 
"a" represents the western arms,
"b" the cavities, 
"c" the arc-like structure,
and "d" the eastern arms. 
The artifact region is shown by the hatched ellipse. 
The directions of the spider pattern are indicated by dotted lines. 
"e" represents the local peak in the logarithmic image of the FS Tau
binary at lower left of the top panel. 
The positions of the FS Tau primary and secondary stars are indicated by 
the big and small cross symbols, or "Aa" and "Ab," respectively. 
The cone-shaped reflection nebula R1 extends to $8\arcsec$~ northeast from 
its illuminating source Haro 6-5 B. 
The dashed lines in the top panel indicate the straight edge of the dark i
cloud suggested by Krist et al. (1998).  
The FOVs are 36\farcs3$\times$49\farcs4 in the top panel 
and 18\farcs8$\times$22\farcs6 in the bottom panel. 
North is up and east is to the left.} 
\label{HST}
\end{center}
\end{figure}

Figure \ref{HST} represents the F606W-band Stokes $I$ images of the 
FS Tau/Haro 6-5 B field and the closeup FS Tau binary. 
The primary and secondary stars of the FS Tau binary were resolved 
in these images.
The projected separation and P.A. were 0\farcs23$\pm$0\farcs01 and 
104\fdg69$\pm$6\fdg0, respectively. 
The F606W-band image reveals the western arms ("a" in Fig. \ref{HST}), 
the northeast and southwest cavities ("b" in Fig. \ref{HST}), 
and the arc-like structure ("c" in Fig. \ref{HST}) 
in common with the $H$-band coronagraphic image. 
Also, we discovered two eastern arms ("d" in Fig. \ref{HST}) 
in the F606W-band image. 
One is sharp, but the other is not. 
These western and eastern arms cover the southwest and northeast 
cavities with opening angles of $\sim$45\degree~ and $\sim$40\degree~, 
respectively. 

A local peak was identified at 1\farcs0 (140 AU) from the 
binary in the F606W-band images before and after the PSF-subtraction 
("e" in Fig. \ref{HST}). 
The FWHM was $\sim$0\farcs8 (110 AU). 
We identified the local peak as an extended dusty structure, 
not a stellar object (\S \ref{ARM2}). 
Its F606W-band magnitude was measured to be 16.0$\pm$0.5 mag arcsec$^{-2}$
with a 0\farcs4 radius aperture;
however, it is located in the artifact region and on the diffraction pattern 
of the spider.

To evaluate the PSF-subtraction in the $HST$ image, 
a pseudo-PSF binary was created using a dwarf, 
2MASS J07040548-0350493 ($HST$ program GO 9588 proposed by H. Bond). 
The flux ratio of the pseudo-binary was set to that of the FS Tau binary 
(20.8; Krist et al. 1998). 
The projected separation of the pseudo-binary had the same value as 
the FS Tau binary in the F606W-band image. 
We subtracted the PSFs created by the Tiny Tim from the pseudo-binary. 
The PSF-subtraction produced artifacts within about 1\farcs6 
from the central star (Fig. \ref{HST}). 
The limiting magnitude of the observations was calculated
from standard deviation of the sky regions,
which was set far from the central star and the nebulosity.
We derived that the limiting magnitude with S/N=3 was 22.1 mag arcsec$^{-2}$.

\subsection{F606W-band Polarimetric Image}

\begin{figure}
\begin{center}
\FigureFile(120mm,210mm){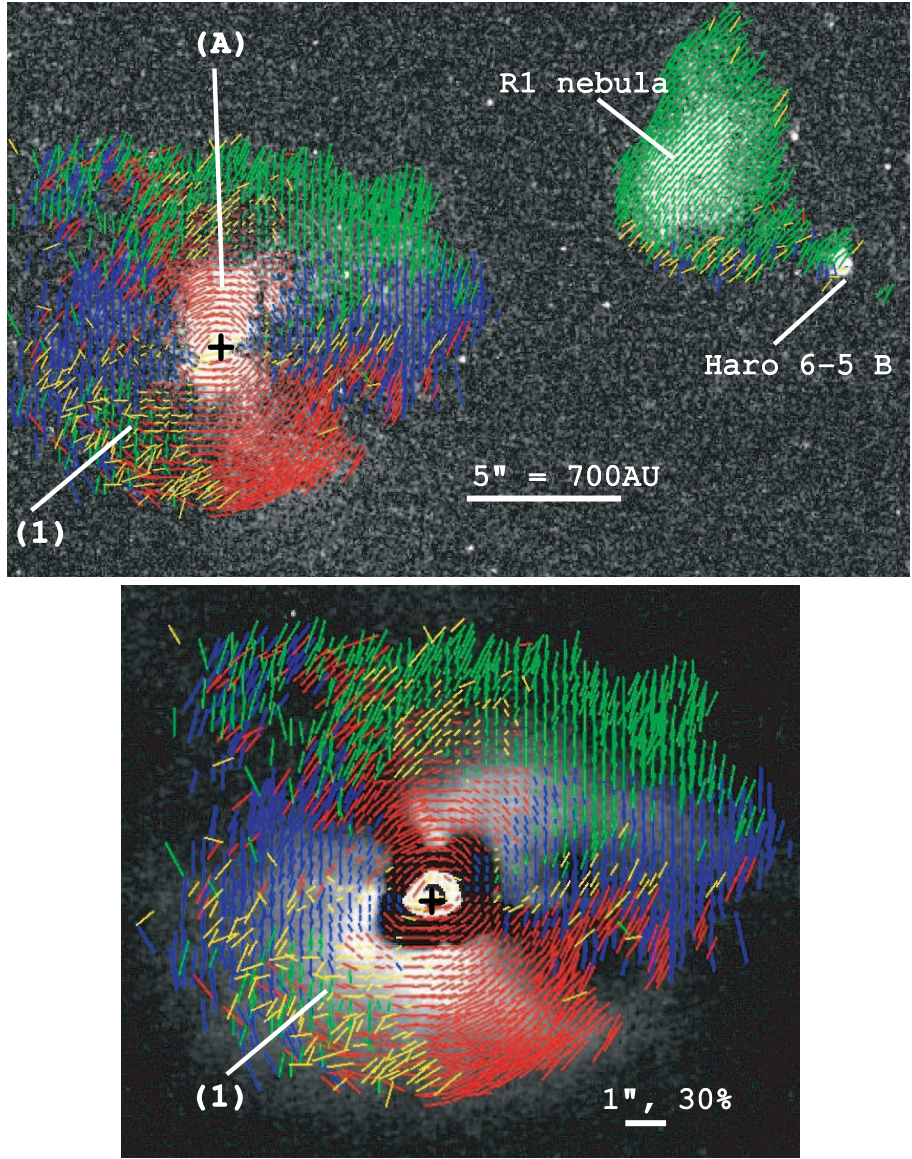}
\caption{The F606W-band polarized intensity image ($top$) and 
$H$-band coronagraphic image ($bottom$; same as Fig. \ref{CIAO}) 
with the F606W-band polarization vectors. 
The vectors are 0\farcs25$\times$0\farcs25 binned polarization vectors 
and are delineated by intensities above $3\sigma$ in each polarized image. 
The vectors are also classified as follows: 
$\theta_{\mathrm{centroA}} < 30\degree$ and 
$\theta_{\mathrm{centroB}} > 30\degree$ ($red$), 
$\theta_{\mathrm{centroA}} < 30\degree$ and 
$\theta_{\mathrm{centroB}} < 30\degree$ ($blue$), 
$\theta_{\mathrm{centroA}} > 30\degree$ and 
$\theta_{\mathrm{centroB}} < 30\degree$ ($green$), 
$\theta_{\mathrm{centroA}} > 30\degree$ and 
$\theta_{\mathrm{centroB}} > 30\degree$ ($yellow$). 
$\theta_{\mathrm{centroA}}$ and $\theta_{\mathrm{centroB}}$ 
are defined as the deviation from the centro-symmetric polarization patterns 
with respect to the FS Tau binary and Haro 6-5 B, respectively (see text). 
The polarization patterns are broadly centro-symmetric around the 
circumbinary disk, resulting from illumination by the FS Tau binary. 
"A" represents the local peak in the disk. 
The polarization vector is non-centro-symmetric around the $H$-band bright 
portion of the disk ("1"). 
The FOV is 19\farcs0$\times$30\farcs6 in the top image and 
13\farcs8$\times$16\farcs5 in the bottom image. 
North is up and east is to the left.} 
\label{irowake}
\end{center}
\end{figure}

Figure \ref{irowake} represents the F606W-band polarized intensity image
and $H$-band coronagraphic image, together with polarization vectors. 
To clarify the sources of the polarizations around the FS Tau binary, 
the polarization angles were investigated. 
We defined the deviation from the centro-symmetric polarization angle 
centered on the binary and Haro 6-5 B as 
$\theta_{\mathrm{centroA}} 
= \vert \theta - \mathrm{P.A._{\mathrm{A} \bot}} \vert$,
and 
$\theta_{\mathrm{centroB}} 
= \vert \theta - \mathrm{P.A._{\mathrm{B} \bot}} \vert$. 
These angles represent the observed polarization angles; 
$\mathrm{P.A._{\mathrm{A} \bot}}$ and $\mathrm{P.A._{\mathrm{B} \bot}}$ 
are perpendicular to the P.A. with respect to the FS Tau primary and 
Haro 6-5 B, respectively, at each position. 
The red-painted polarization vectors in Figure \ref{irowake} represent 
$\theta_{\mathrm{centroA}} < 30\degree$~ and 
$\theta_{\mathrm{centroB}} > 30\degree$~; 
the blue-painted polarization vectors 
$\theta_{\mathrm{centroA}} < 30\degree$~ and 
$\theta_{\mathrm{centroB}} < 30\degree$~; 
the green-painted polarization vectors 
$\theta_{\mathrm{centroA}} > 30\degree$~ and 
$\theta_{\mathrm{centroB}} < 30\degree$~; 
the yellow-painted polarization vectors 
$\theta_{\mathrm{centroA}} > 30\degree$~ and 
$\theta_{\mathrm{centroB}} > 30\degree$~. 
If the FS Tau binary were the only illuminating source in the FS Tau/Haro 6-5 B 
field, the angles $\theta_{\mathrm{centroA}}$ would fall below 30\degree~ 
at all points.
This means that the blue-painted polarization vectors would be present 
between the P.A. = 57\degree~ to 135\degree~ and 250\degree~ to 303\degree~
if the binary had a geometrically flat circumbinary disk with 
a major axis along a P.A. = 30\degree~ and an inclination by 35\degree~
(\S \ref{DISK}). 
The red-painted polarization vectors would be present at the other P.A.s. 
Their degrees of polarization depend on the scattering angles. 

The FS Tau binary system has two polarimetric characteristics: 
small $\theta_{\mathrm{centroA}}$ features at the local peak 
($\theta_{\mathrm{centroA}} = 5\degree \pm 4\degree$), 
the southwest arm ($7\degree \pm 5\degree$), 
and the western cavity ($13\degree \pm 11\degree$); 
and large $\theta_{\mathrm{centroA}}$ features at the portion of the
southeast that is bright in the $H$-band (30\degree~$\pm$22\degree~) 
and at the north side 
at a distance from the binary, $r$, of $r>$ 420 AU  
(53\degree$\pm$17\degree~). 
Causes of these large deviations in the polarization angles 
will be discussed in the next section. 
The polarized image does not show parallel polarization 
vectors or null points near the central binary both of which are 
indicative of a polarization disk. 
The averaged F606W-band polarization was $P = 11 \pm 5$\% in $r < 560$ AU, 
which is consistent with the result of Gledhill \& Scarrott 
(10\% at the $R$-band). 

\section{Discussion}

\subsection{Circumbinary Disk}

We consider that the emission nebulae are mostly attributed to the
circumbinary disk.

\subsubsection{Optical and NIR Imaging/Polarimetric Characteristics of the Disk}
\label{DISK}

\begin{figure}
\begin{center}
\FigureFile(140mm,270mm){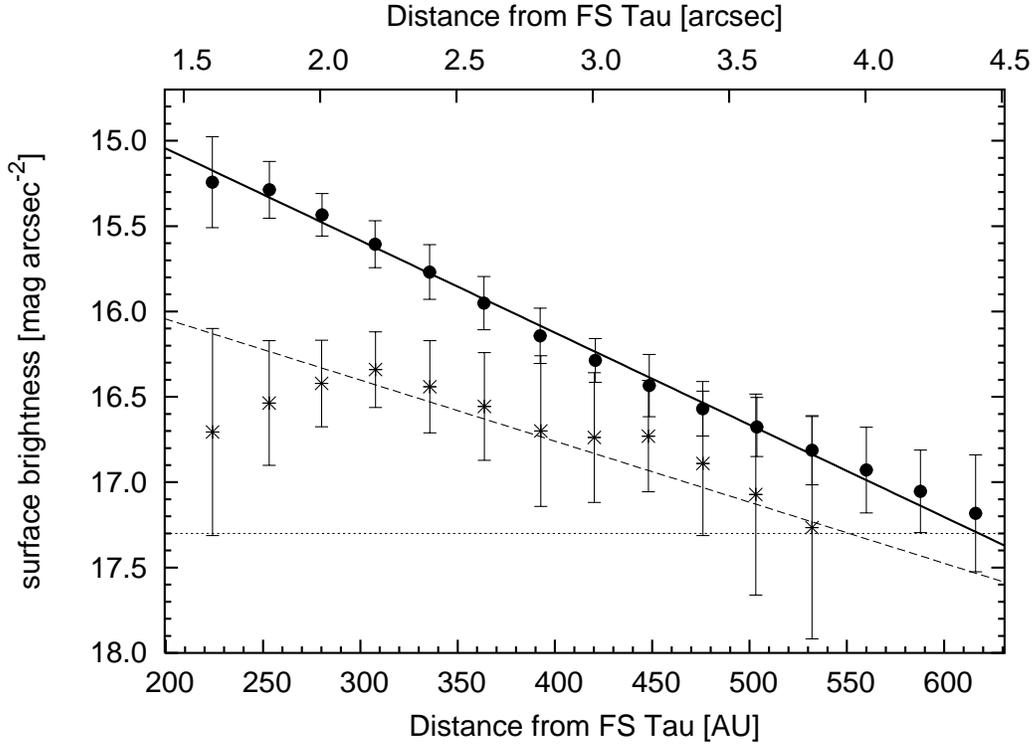}
\caption{The $H$-band surface brightness profile of the southeast 
(filled circle; P.A. = 93\degree~ to 175\degree~) and northwest 
(asterisk; P.A. = 280\degree~ to 355\degree~) sides of the binary. 
The brightnesses are averaged in a 0\farcs2$\times$0\farcs2 aperture at 
$r = 220 - 630$ AU. 
The error bars represent standard deviations. 
The solid line represents a $r^{-1.9 \pm 0.1}$ power-law, 
and the dashed line represents a $r^{-1.5 \pm 0.3}$ power-law. 
The limiting magnitude of the observations (17.3 mag arcsec$^{-2}$) 
is shown by the horizontal dotted line.} 
\label{RP}
\end{center}
\end{figure}

Figure \ref{RP} shows the $H$-band radial profile along the bright 
southeast region. 
The $H$-band brightness was 15.2$\pm$0.3 mag arcsec$^{-2}$ at 
$r$ = 220 AU and 16.8$\pm$0.2 mag arcsec$^{-2}$ at $r$ = 530 AU.
We fit a single power-law of $r^{-1.9 \pm 0.1}$ to the southeast region.
Its slope was shallower than that of circumstellar disks around other 
classical T Tauri stars, such as FN Tau (-2.5; Kudo et al. 2008), 
TW Hya (-2.6; Weinberger et al. 2002), and GM Aur 
(-3.5; Schneider et al. 2003), in the $H$-band. 
The surface brightness of a geometrically flat and optically thick structure
fits a single power-law of $r^{-3}$, 
whereas the surface brightness of a flared structure 
with well-mixed gas and dust fits a power law of $r^{-2}$ 
(Whitney \& Hartmann 1992). 
We argue that the southeast region corresponds to a part of the 
circumbinary disk and is more flared than the other T Tauri disks. 
One may consider that the structure is a flattened envelope rather than
a disk.
But the central binary is not heavily reddened.
The combined near-infrared colors are $J-H=1.46\pm0.04$ mag
and $H-K=1.07\pm0.03$ mag,
corresponding to the intrinsic colors of classical
T Tauri stars (Meyer et al. 1997) with a moderate extinction
($A_{\rm V} \sim 6$ mag).
Based on these colors, we consider that the structure is a circumbinary disk.
The $H$-band radial profile of the northwest region of the binary
are also shown in Figure 4.
It is fainter than the southeast region at all distances.
We fit a single power-law of $r^{-1.5\pm0.3}$ to the northwest region.
Its slope seems shallower than that of the southeast region,
but its faintness prevents us from further discussion.

\begin{figure}
\begin{center}
\FigureFile(140mm,270mm){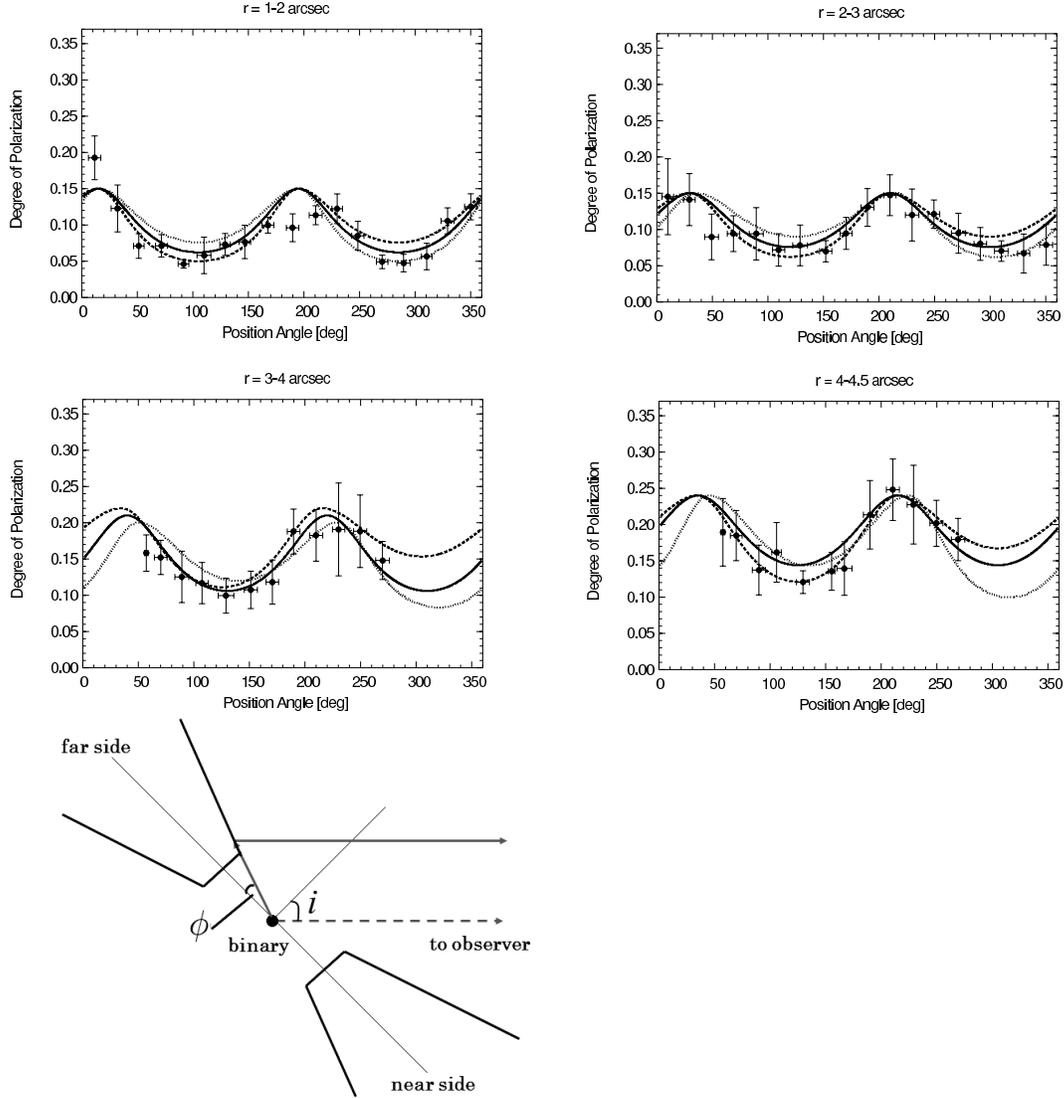}
\caption{The azimuthal polarization profiles for various distances 
from the binary, starting from the top left ($r$ = $1 - 2\arcsec$~) 
and ending at the bottom right ($r$ = $4 - 4.5\arcsec$~). 
The observed polarizations with $\theta_{\mathrm{centroA}} < 30\degree$ 
are applied to exclude the polarization due to multiple 
scattering events in the circumbinary disk. 
The observed polarizations in the $r > 3\arcsec$~ north 
(P.A. = 280\degree~ to 55\degree~) are not shown due to the contamination of 
the light from Haro 6-5 B. 
The solid curves represent the best match of the flat ($\phi = 0\degree$) 
circumbinary disk model. 
The dashed curves represent the best match of the flared ($\phi = 5\degree$) 
disk, where the southeast side corresponds to the near side. 
The dotted curves are the best match of the flared ($\phi = 5\degree$) disks, 
where the northeast side is the near side. 
These $\chi^{2}$ are shown in Table 1. 
"$\xi$" is the P.A. of the disk major axis, 
which is perpendicular to the plane of the bottom left panel. 
"$i$" is the disk inclination, and $\phi$ is the flared angle. 
No data are shown at $r < 1\arcsec$~ due to the large uncertainty. } 
\label{PAVSPOLARI}
\end{center}
\end{figure}

Figure \ref{PAVSPOLARI} represents the F606W-band azimuthal 
profiles of the degrees of polarization around the FS Tau binary. 
The degrees of polarization were derived from a 0\farcs25$\times$0\farcs25 
aperture.
We determined that the outer radius of the disk is consistent with 
that of the southeast region (4\farcs5 = 630 AU), which is bright in the
$H$-band.
The observed polarizations of $\theta_{\mathrm{centroA}} > 30\degree$ 
(i.e., the green- and yellow-painted vectors in Fig. \ref{irowake}) 
were not applied in order to exclude the polarizations caused 
by multiple scattering events from dust grains
in the southeast region (\S \ref{MULTI}). 
In addition, the polarizations in the $r >$ 3\arcsec~ north 
(P.A. = 280\degree~ to 55\degree~) were masked due to the contamination 
by light from the cavity wall associated with Haro 6-5 B (\S \ref{BRIGHT}). 
The maximum degrees of polarization, $P_{\mathrm{max}}$, 
were observed in the northeast and southwest sides of the disk, 
and the minima, $P_{\mathrm{min}}$, were observed in the southeast 
and northwest sides.  

We constructed a simple circumbinary disk model 
to investigate the polarization by dust grains in the FS Tau binary system.
The FS Tau binary has a prominent smooth emission feature at
10 \micron~ (Furlan et al. 2006), 
which indicates sub-micron-sized amorphous silicate dusts 
in the disk surface layer.
Such dust grains show Rayleigh-like scattering.
The GG Tau binary also shows a strong 10 \micron~ silicate emission 
feature similar to that of the FS Tau binary (Furlan et al. 2006). 
The circumbinary disk around GG Tau displays a high polarization amplitude 
($\sim$50\%), with a centro-symmetric pattern in the NIR wavelengths
(Silber et al. 2000),
which is well reproduced by Rayleigh-like scattering ($a \ll \lambda$) 
from sub-micron dust grains in the disk. 
Therefore, we assume that "single" Rayleigh-like scattering
occurs in the FS Tau circumbinary disk. 
Using small dust grains with a power-law size distribution of 
$n (a) \propto a^{-3.5}$ between $0.005-0.25$\micron~ 
(hereafter MRN dust; Mathis et al. 1977), the degree of polarization, $P$, 
depends on the scattering angle, $\theta_{\mathrm{scat}}$, as below:  
\begin{eqnarray}
P 
= \frac{I_{\bot} - I_{\Vert}}{I_{\bot} + I_{\Vert}} 
= \frac{1 - \cos^2 \theta_{\mathrm{scat}}}{1 + \cos^2 \theta_{\mathrm{scat}}};\\
\theta_{\mathrm{scat}} 
= \cos^{-1}\left( \frac{y \tan (i \pm \phi)}{(x^{2}+y^{2}+y \tan(i \pm \phi))^{1/2}} \right),
\end{eqnarray}
where $I_{\Vert}$ and $I_{\bot}$ are the intensities of scattered light 
in polarization modes parallel and perpendicular to the scattering plane, 
respectively (van de Hulst 1957). 
The symbol $i$ is the disk inclination, $\phi$ is the flared angle of the disk, 
and $x$, $y$ are the coordinates deprojected along the 
major and minor disk axes, 
respectively, with their origin at the center of the primary star 
(Fig. \ref{PAVSPOLARI}), i.e., $+ \phi$ for the near side of the disk along our 
line of sight and $- \phi$ for the far side. 
Three parameters, $i$, $\phi$, and the P.A. of the disk major axis, $\xi$, 
are free parameters. 
These parameters were determined from a $\chi^{2}$ minimization method at 
$r$ = 1\arcsec~ intervals. 

The disk model shows a centro-symmetric polarization pattern around the binary. 
The $P_{\mathrm{max}}$ occurs at two points, each at a 90\degree~ scattering 
angle (i.e., disk major axis).
The disk minor axis corresponds to the positions of the $P_{\mathrm{min}}$, 
where the scattering angles are far from 90\degree~.
If the disk is flared, $P_{\mathrm{min}}$ occurs at the near side of 
the disk.
Notice that a spherical dust distribution, such an envelope, 
does not produce a sinusoidal pattern; 
in that case, the degree of polarization is constant
as a function of the scattering angle (or P.A.). 
Because the observed $P_{\mathrm{max}}$ occurs at 
P.A. = 30\degree~ and 210\degree~
at $r = 2 - 3$\arcsec~ ($280-420$ AU; Fig. \ref{PAVSPOLARI} and Table 1), 
we determined the disk major axis at P.A. = 30\degree~ 
and a disk inclination of 35\degree~. 
The flat disk model ($\phi = 0\degree$) agree well with the observed 
polarization at $r = 2 - 3\arcsec$~. 
The polarization is at a minimum at P.A. = 120\degree~ and 300\degree~. 
Assuming the southeast side is brighter than the northwest side 
in the $H$-band due to the forward scattering of dust, 
the southeast side is located closer to us
($\theta_{\mathrm{scat}} \sim 55\degree$), 
and the northwest side is oriented away from us 
($\theta_{\mathrm{scat}} \sim 125\degree$).
However, the radial profile of the surface brightness of the disk
is not consistent with the flat-disk model.
The $\chi^{2}$ minimization method shows
that the southeast side of the disk has a lower degree of polarization 
than the northwest side at $r = 4 - 4.5\arcsec$~ ($560-630$ AU). 
The flared disk model ($\phi = 5\degree$) possibly explains 
the observed polarization at $r = 4 - 4.5\arcsec$~, 
where the disk is inclined by 30\degree~, with its southeast side toward us.

The P.A. of the disk major axis and the inclination of the disk 
are consistent with those of the binary orbit,
whose major axis is along the P.A. = 18\degree$\pm$7\degree~ 
and an inclination of 20\degree$\pm$5\degree~ (Tamazian et al. 2002). 
The $\chi^{2}$ minimization method also shows 
($\xi$, $i$, $\phi$) = (15\degree~, 40\degree~, 0\degree~) 
at $1-2\arcsec$~ and (35\degree~ to 40\degree~, 30\degree~ to 35\degree~, 
0\degree~ to 5\degree~) at $3-4\arcsec$~. 

In conclusion, the circumbinary disk is characterized by the following 
geometry: the radius of the disk is 630 AU (4\farcs5), 
the disk major axis is along the P.A. = 15\degree~ to 40\degree~, 
and the disk is inclined by 30\degree~ to 40\degree~, 
where the southeast side (P.A. = 105\degree~ to 130\degree~) 
corresponds to the side nearest to us. 
The outer ($>$560 AU) portion of the disk is possibly more flared 
than its inner portion. 

\subsubsection{Southeast to Northwest Brightness Asymmetry}
\label{BRIGHT}

The circumbinary disk shows an unambiguous southeast-northwest asymmetry. 
The brightness ratio of the southeast and northwest sides at $r = 280 - 630$ AU 
is $1.6 \pm 0.4$ in the $H$-band. 
We suggest that anisotropic scattering of the dust causes the asymmetric 
brightness of the disk and that the southeast side is brighter 
due to forward scattering of the dust. 
This indicates that the southeast side corresponds to the near side. 
Based on this hypothesis, the brightness ratio 
($B_\mathrm{near}/B_\mathrm{far}$) of the disk is described as: 

\begin{eqnarray}
\frac{B_\mathrm{near}}{B_\mathrm{far}} 
&=& \left( \frac{1 + g^{2} - 2 g \cos \bar{\theta}_{\mathrm{Nscat}}}{1 + g^{2} - 2 g \cos \bar{\theta}_{\mathrm{Fscat}}} \right)^{-3/2},
\end{eqnarray}
where $g$ is the asymmetry parameter of the Henyey-Greenstein 
scattering phase function (Henyey \& Greenstein 1941), 
$\bar{\theta}_{\mathrm{Nscat}}$ is the averaged scattering angle of the near 
side, and $\bar{\theta}_{\mathrm{Fscat}}$ is that of the far side (Eq. 7). 
For the disk parameters in \S \ref{DISK} ($r = 630$ AU, $\xi =$ 15\degree~
to 40\degree~,
$i =$ 30\degree~ to 40\degree~, and $\phi =$ 0\degree~ to 5\degree~), 
the $H$-band brightness ratio was fit with $g \sim 0.2$. 
This value is consistent with that of the MRN dust ($g \sim 0.1$) 
shown by Kim et al. (1994), who assumed single scattering by the dust 
in a diffuse interstellar cloud ($R_{\mathrm{V}} = 3.1$). 
For the FS Tau circumbinary disk, such small dust grains probably produce 
single-scattered light, as the disk is optically thin in 
the $H$-band (\S \ref{MULTI}).

\subsubsection{Northwest Side of the Binary}
\label{blue}

The northwest side of the circumbinary disk are bright in the optical
wavelengths but faint in the near-infrared wavelengths (Fig. 6).
We measured its color.
Within 1\farcs5 from the central star, the PSF subtractions are
not perfect either in the optical image nor in the near-infrared image.
Beyond 4\arcsec~ from the central star the disk is very faint
in the near-infrared wavelengths.
We measured the surface brightness of the disk in the region between
1\farcs5 and 4\farcs0 from the central star and 
the P.A. between 280\degree~ and 355\degree~.
The F606W$- H$ color in this region was derived to be between 1.7 mag and 
3.0 mag. 
We also derived the intrinsic colors, (F606W$- H)_{\mathrm{int}}$, 
of the photosphere of the binary and Haro 6-5 B as $2.44\pm0.02$ mag 
and $2.00\pm0.02$ mag, 
respectively, using the relationship (F606W$- H)_{\mathrm{int}} 
= (V - H)_{\mathrm{int}} - (0.477\pm0.013)(V - R)_{\mathrm{int}} 
+ (0.001\pm0.005)(V - R)_{\mathrm{int}}^2$ (Sirianni et al. 2005). 
The intrinsic colors, $(V - R)_{\mathrm{int}}$ and $(V - H)_{\mathrm{int}}$, 
were derived from those of main sequence stars (Ducati et al. 2001) 
whose spectral types were the same as the FS Tau primary 
(M0; Hartigan \& Kenyon 2003) and Haro 6-5 B (K5; White \& Hillenbrand 2004). 
Comparing the disk color with the colors of the point sources,
we cannot identify whether the disk is illuminated by the FS Tau binary
or Haro 6-5 B.

On the other hand, the polarization map clearly indicates
the contribution of Haro 6-5 B.
We claim that the scattered light and optical emission lines from 
Haro 6-5 B change the degree of polarization and polarization angle 
on the north side at $r > 3\arcsec$~. 
Several observations identified a cavity wall in the north side, 
which is created by the blue-shifted outflow from Haro 6-5 B 
(Gledhill \& Scarrott 1989; Eisl$\ddot{\mathrm{o}}$ffel \& Mundt 1998). 
Gledhill \& Scarrott (1989) reported that the cavity wall has 
non-centro-symmetric polarization 
patterns with respect to the FS Tau binary, bluer colors than those of 
the south side of the binary, and an H$_{\alpha}$ emission line. 
We confirm the radiation from Haro 6-5 B at $r > 3\arcsec$~ north 
(P.A. = 280\degree~ to 55\degree~) in the F606W-band polarimetric image. 
The green-painted polarization vectors are predominant (56\% of the total) 
in the $r > 3\arcsec$~ north side (Fig. \ref{irowake}). 
The averaged $\theta_{\mathrm{centroA}}$ is $53\degree \pm 17\degree$, 
whereas the averaged $\theta_{\mathrm{centroB}}$ is $11\degree \pm 8\degree$. 
The averaged $P$ is $16 \pm 7\%$. 
These $\theta_{\mathrm{centroB}}$ and $P$ are comparable to those of 
the cone-shaped reflection nebula (hereafter R1 nebula) that 
extends 8\arcsec~ northeast from its illuminating source Haro 6-5 B. 
In addition, the radiation from Haro 6-5 B is a possible cause of the 
blue-painted polarization vectors at $r > 3\arcsec$~ north. 
The averaged $\theta_{\mathrm{centroA}}$ is slightly larger than the 
averaged $\theta_{\mathrm{centroB}}$ ($20\degree \pm 7\degree$ vs. 
$14\degree \pm 9\degree$). 
We argue that the cavity wall blocks the north 
side of the circumbinary disk at $r > 3\arcsec$~.
On the other hand, for the north side at $ r < 3\arcsec$~, the averaged 
$\theta_{\mathrm{centroA}}$ is $13\degree \pm 13\degree$ and the averaged 
$\theta_{\mathrm{centroB}}$ is $44\degree \pm 24\degree$,
indicating the scattered light from the FS Tau binary.

\subsubsection{Southeast Side of the Disk}
\label{MULTI}

\begin{figure}
\begin{center}
\FigureFile(140mm,270mm){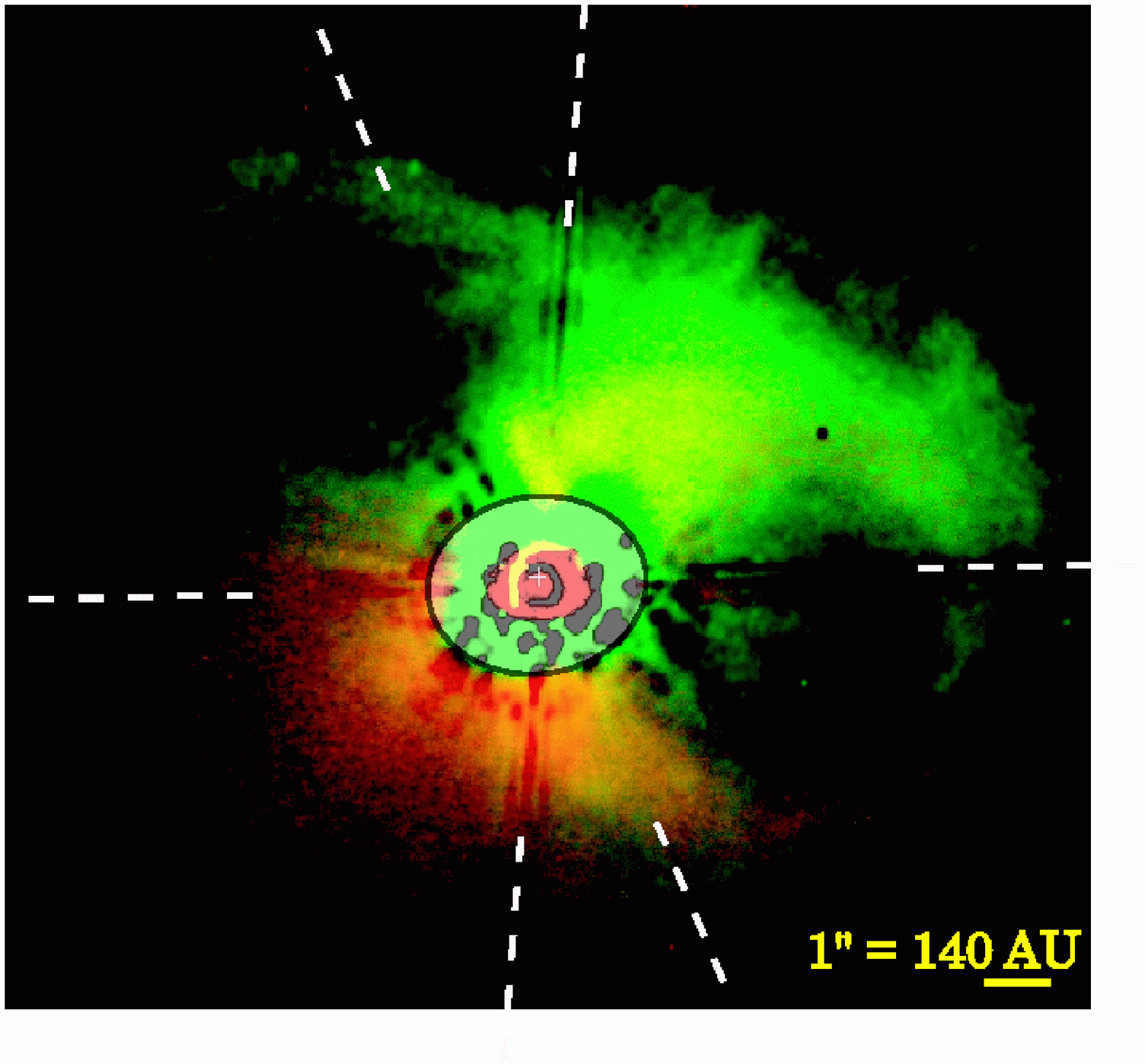}
\caption{The F606W$- H$ color image of the FS Tau binary. 
The hatched ellipse marks the artifact region caused by the PSF-subtraction 
in both the F606W-band ($green$) and $H$-band images ($red$). 
The directions of the spider pattern are indicated by dashed lines. 
The FOV is 16\farcs5$\times$17\farcs5. 
North is up and east is to the left.} 
\label{V-H}
\end{center}
\end{figure}

The southeast side of the circumbinary disk is bright in the near-infrared
wavelengths but faint in the optical wavelengths (Fig. 6).
We measured its color in the region between 1\farcs5 and 4\farcs5 from
the central star and the P.A. between 93\degree~ and 175\degree~.
The F606W$-H$ color of this region of the disk is derived to be
4.2$\pm$0.2 mag, 
which is redder than the intrinsic color of the FS Tau binary (Fig. \ref{V-H}). 
Figure \ref{irowake} displays
a weak centro-symmetric ($\theta_{\mathrm{centro}}\sim 30\degree$) 
polarization pattern around the southeast portion, 
despite the high signal to noise ratio ($\geq 6$). 
We consider that the southeast portion corresponds to a part of the 
circumbinary disk. 

Given the (F606W$- H)_{\mathrm{int}}$ of the binary ($2.44\pm0.02$ mag) and the 
F606W$ - H$ color of the southeast portion (4.2$\pm$0.2 mag), 
the color excess of the southeast portion is $E ($F606W$ - H) = 1.8\pm0.2$ mag. 
We assume that extinction by dust grains around the binary reddens 
the southeast portion. 
This extinction, $A_{\lambda}$, was derived as $2.3\pm0.3$ mag 
in the F606W-band and $0.5\pm0.1$ mag at the $H$-band. 
We estimated an optical depth, $\tau_{\lambda}$, of $2.1\pm0.3$ in the 
F606W-band and $0.5\pm0.1$ at the $H$-band, using the relationship 
$A_{\lambda} = 1.086 \tau_{\lambda}$ (e.g., Stark et al. 2006). 
Thus, the southeast portion is optically thick in the F606W-band, 
and some of the scattered light comes from the southeast portion,
where it expected multiple scattering events from dust grains.
Further circular polarization observations are needed to perceive 
the multiple scattered light events.

\subsection{Outflow}
\label{outflow}

No optical jet was detected within a few arcseconds of the FS Tau binary 
(Woitas et al. 2002). 
Although $^{13}$CO emission was observed along the southeast side, 
it is unknown whether this emission arises from the diffuse wind 
from the binary (Dutrey et al. 1996). 
The circumbinary disk model (\S \ref{DISK}) predicts a red-shifted outflow 
driven by the primary or secondary stars along the southeast direction 
(P.A. = 105\degree~ to 135\degree~), 
given that the circumstellar disks are coplanar with the circumbinary disk, 
and the outflow axis is perpendicular to the circumstellar disk plane. 

In contrast, both the $H$-band and F606W-band images show evidence
of other outflows: the western and eastern arms + cavity systems, 
and the arc-like structure (Figures \ref{CIAO} and \ref{HST}).
The arc-like structure located at $r \sim 900$ AU west is seen faintly
in the $H$-band image, but is clearly detected at above $10 \sigma$ 
in the F606W-band image. 
This structure appears to connect the western arms or to have
an appearance similar to bow shocks excited by Herbig-Haro flows. 
The two western arms encompass the southwest cavity with an opening 
angle of $\sim$45\degree~. 
The eastern arms + cavity system, only detected in the F606W-band image,
plausibly mixes with the cavity wall associated with Haro 6-5 B (\S \ref{blue}).
We argue that these arms + cavity systems are similar to the features 
created by bipolar outflows in several T Tauri systems 
(e.g., Haro 6-5 B: Eisl$\ddot{\mathrm{o}}$ffel \& Mundt 1998, 
HL Tau; Lucas et al. 2004). 
We measured surface brightness of the cavities.
The regions of the cavities were defined as circular areas
with 3\farcs5 radius centered at 12\farcs4 with the P.A. of 43\fdg5 
from the central star
and centered at 9\farcs9 with P.A=245\fdg7
for the northeast cavity and the southwest cavity, respectively.
The (b) marks of the right figure of Fig. 2 correspond to the central positions
of the cavity regions.
The southwest and northeast cavities have comparable F606W-band 
brightnesses of $21.2\pm0.2$ mag arcsec$^{-2}$, 
whereas their $H$-band brightness 
are lower than the limiting magnitude (17.3 mag arcsec$^{-2}$). 
Their F606W$-H$ colors are $< 3.9$ mag i.e., $\tau_{\mathrm{F606W}} < 1.9$ 
and $\tau_{\mathrm{H}} < 0.4$ (see \S \ref{MULTI}), 
assuming that the cavities are reddened by dust extinction. 
The outflow from the binary would sweep up materials in the cavities. 
The polarization of the northeast cavity is $18 \pm 5\%$, 
and that of the southwest cavity is $19 \pm 6\%$. 
These polarizations are relatively higher than the other structures located 
within 560 AU of the binary (11\%). 
The southwest cavity shows 
$\theta_{\mathrm{centroA}} = 13\degree \pm 11\degree$, 
indicating the efficiently scattered light from the binary. 
The northeast cavity has
$\theta_{\mathrm{centroA}} = 22\degree \pm 15\degree$ and 
$\theta_{\mathrm{centroB}} = 22\degree \pm 10\degree$. 
The northeast cavity is illuminated by mixed radiation from the binary 
and Haro 6-5 B, although the dereddened color of the mixed radiation is unclear.
We claim that the outflow emanates to the P.A. = 220\degree~ to  250\degree~, 
which is equal to the axisymmetric orientation of the western arms. 
This direction is not consistent with that calculated by our disk model 
(P.A. = 105\degree~ to 135\degree~ or 285\degree~ to 315\degree~). 
These differences are possibly produced by the mechanisms discussed in the 
following sections. 

\subsubsection{Disk Misalignment}

One possible mechanism is that the primary/secondary circumstellar disks 
are not coplanar with the circumbinary disk. 
For T Tauri binaries separated by several AU (e.g., DQ Tau; Mathieu et al. 1997,
UZ Tau E; Prato et al. 2002) and T Tauri binaries with moderate separation
(e.g., GG Tau; Tamazian et al. 2002), the circumbinary disks are coplanar 
with each binary orbital plane. 
In contrast, for widely separated T Tauri binary systems,
$K$-band polarimetric observations suggest that circumstellar 
disks are not perfectly coplanar but differ slightly from each other 
($\sim 20\degree$; Jensen et al. 2004). 

Bate et al. (2000) demonstrated that during about 2000 binary orbital 
periods circumstellar disks are roughly aligned with a binary orbital 
plane by tidal torques. 
This timescale is estimated to be $5.5 \times 10^{5}$ yrs for the FS Tau 
binary, given its separation of 39 AU and its total binary mass of 0.78 \MO~ 
(Tamazian et al. 2002). 
We suggest, based on a comparison of the alignment timescale with the 
stellar age based on the shape of the SED ($10^{5}-10^{6}$ yrs; 
Andrews \& Williams 2005),
that the circumbinary disk is still misaligned with the circumstellar disks.

\subsubsection{Gravitational Instability and Inhomogeneous Distributions of the Disk Materials}
\label{ARM2}

Another possibility is that the western and eastern arms are formed 
by other mechanisms, such as gravitational disk instability 
or inhomogeneous distributions of the circumbinary materials. 

A circumstellar disk associated with AB Aur has spiral arms that 
are expected to be formed by the gravitational instability of the disk 
(Fukagawa et al. 2004). 
Nelson et al. (1998) revealed that such a spiral arm appears in a 
circumstellar disk if $Q \leq 2.0$, where $Q = c_{s}\Omega/(\pi G \Sigma)$ 
is Toomre's $Q$-value, $c_{s}$ is the sound speed, $\Omega$ 
is the angular velocity, and $\Sigma$ is the surface density of the disk. 
However, the FS Tau circumbinary disk is less massive (0.002 \MO~; 
Andrews \& Williams 2005). 
Its Toomre's $Q$-value is much larger than 2.0, given the disk radius of 
630 AU, the surface density distribution of $\Sigma \propto r^{-0.5}$ 
(Kitamura et al. 2002), and $c_{s} = 190$ m s$^{-1}$ 
($T = 10$ K, for typical T Tauri disks). 
Thus, we conclude that gravitational disk instability does not occur 
in the circumbinary disk. 

The second possible mechanism is that a companion shapes the
spiral arm-like features (e.g., Bate et al. 2003). 
However, high spatial resolution observations have detected no other 
companion around the FS Tau binary (Chen et al. 1990; Simon et al. 1992; 
Krist et al. 1998; Hartigan \& Kenyon 2003; Connelley et al. 2009). 
We identified a local peak at $r \sim 140$ AU north in the F606W-band 
polarimetric images ("e" in Fig. \ref{HST} and "A" in Fig \ref{irowake}). 
This structure corresponds to the "small arc" detected by Krist et al. (1998). 
This local peak is extended (its FWHM of 110 AU) and shows $P = 15$\% 
and $\theta_{\mathrm{centroA}} = 5\degree$. 
These indicate that the local peak is an extended dusty structure 
associated with the binary, not a stellar companion. 
The lower limit of the local peak mass is estimated to be 
$\sim 0.3$ $M_{\oplus}$ from its F606W-band brightness (16.0 mag arcsec$^{-2}$).
Such a low mass structure cannot create the disk instability. 

Finally, several studies have demonstrated arm-like structures as 
resulting from inhomogeneous distributions of materials in a circumbinary disk 
(G$\ddot{\mathrm{u}}$nther \& Kley 2002; Ochi et al. 2005). 
The observed local peak could be such an inhomogeneity. 
However, no inhomogeneous gas and dust structures around the binary are
detected in the millimeter wavelength observations due to insufficient 
spatial resolution ($\sim 3\arcsec$~) and flux sensitivity 
(Dutrey et al. 1996; Andrews \& Williams 2005). 
The Atacama Large Millimeter/submillimeter Array (ALMA) will clarify 
whether the arms + cavity systems correspond to outflowing materials 
or to a part of the Keplerian circumbinary disk. 

\section{Conclusions}

Using the Subaru/CIAO, we obtained an $H$-band coronagraphic 
image of the T Tauri binary system FS Tau. 
Combined with the $HST$/ACS (F606W) polarimetric images, 
several features around the binary were identified. 

The circumbinary disk extends to 630 AU in radius. 
The disk is inclined by 30\degree~ to 40\degree~, 
such that the southeast side is nearest. 
The outer portion of the disk may be more flared than the inner portion. 
The disk also displays an unambiguous southeast-northwest brightness asymmetry. 
The southeast side of the disk is significantly red (F606W$-H = 4.2 \pm 0.2$ 
mag). 
It shows a weak centro-symmetric ($\theta_{\mathrm{centroA}} \sim 30\degree$)
polarization pattern in the F606W-band. 
These indicate multiple scattering events
in the optically thick southeast side at F606W-band wavelengths. 
The northeast side of the binary is about neutral in color
(F606W$-H = 1.7$ mag or 3.0 mag)
compared with the FS Tau binary and Haro 6-5 B. 
We argue that the northwest side of the binary at $r > 420$ AU (3\arcsec~) 
is blocked by the cavity wall associated with Haro 6-5 B. 
This idea is supported by the observed polarization angles 
($\theta_{\mathrm{centroA}} = 53\degree \pm 17\degree$ and 
$\theta_{\mathrm{centroB}} = 11\degree \pm 8\degree$). 
The $r < 420$ AU northwest side has 
$\theta_{\mathrm{centroA}} = 13\degree \pm 13\degree$ and 
$\theta_{\mathrm{centroB}} = 44\degree \pm 24\degree$, 
indicating scattered light from the FS Tau binary. 

The western cavity encompassed by the two arm structures is detected 
in both the F606W-band and $H$-band images. 
The eastern cavity, with additional arms, is seen in the F606W-band image, 
but only the cavity appears in the $H$-band. 
We argue that these arms + cavity systems are created by the bipolar 
outflow from the binary along the P.A. = 220\degree~ to 250\degree~. 
Another interpretation is that these structures correspond to 
inhomogeneous distributions of materials in the disk. 
Further high spatial and velocity resolution observations at sub-millimeter 
and millimeter wavelengths (e.g., ALMA) are required to clarify the mechanisms 
of these structures. 

\bigskip

We thank the telescope staff members and operators at the Subaru Telescope. 
We are grateful for fruitful discussions with Tadashi Mukai. 
The $HST$ data presented in this paper were obtained from the 
Multimission Archive at the Space Telescope Science Institute (MAST). 
This study was partly supported by the Global Centers of Excellence (GCOE) 
Program: "Foundation of International Center for Planetary Science" 
from the Ministry of Education, Culture, Sport, Science, and Technology (MEXT). 
T. H. is financially supported by the Japan Society for the Promotion of 
Science (JSPS) for Young Scientists.

\clearpage

\clearpage

\clearpage

\clearpage

\clearpage

\clearpage

\clearpage

\begin{table}[htbp]
\begin{center}
\caption{Best fit parameters for the circumbinary disk. 
$\xi$: the P.A. of the disk major axis; $i$: the disk inclination; 
$\phi$: the flared angle.}
\begin{tabular}{ccccc} \hline\hline
Distance & $\xi$ & $i$ & $\phi$ & $\chi^{2}$ \\\
[arcsec] &[deg]  & [deg] & [deg] &  \\\hline
\multicolumn{5}{c}{Flat Disk} \\\hline
$1-2$   & 15 & 40 & 0 & 1.1 \\
$2-3$   & 30 & 35 & 0 & 2.1 \\
$3-4$   & 40 & 35 & 0 & 1.7 \\
$4-4.5$ & 35 & 30 & 0 & 0.9 \\\hline
\multicolumn{5}{c}{Flared Disk (SE)$^{*}$} \\\hline
$1-2$   & 15 & 40 & 5 & 1.2 \\
$2-3$   & 30 & 35 & 5 & 2.3 \\
$3-4$   & 35 & 30 & 5 & 1.7 \\
$4-4.5$ & 35 & 30 & 5 & 0.8 \\\hline
\multicolumn{5}{c}{Flared Disk (NW)$^{*}$} \\\hline
$1-2$   & 15 & 40 & 5 & 1.5 \\
$2-3$   & 35 & 35 & 5 & 2.4 \\
$3-4$   & 50 & 35 & 5 & 2.0 \\
$4-4.5$ & 45 & 35 & 5 & 1.0 \\\hline
\end{tabular}
\end{center}
$^{*}$ The near side of the flared disk corresponds to the southeast (SE) 
or northwest (NW) sides. \\
\label{Ptable}
\end{table}


\begin{thebibliography}{}

\bibitem[]{} Acke, B., Min, M., van den Ancker, M. E., Bouwman, J., Ochsendorf, B., Juhasz, A., \& Waters, L. B. F. M. 2009, \aap, 502, L17

\bibitem[]{} Armitage, P. J., Clarke, C. J., \& Tout, C. A. 1999, \mnras, 304, 425

\bibitem[Artymowicz \& Lubow(1996)]{al94} Artymowicz, P.,
	\& Lubow, S. H.	1994, \apj, 421, 651
        
\bibitem[Artymowicz \& Lubow(1996)]{al96} Artymowicz, P.,
	\& Lubow, S. H.	1996, \apj, 467, L77

\bibitem[]{}  Andrews, S. M. \& Williams, J. P. 2005, \apj, 631, 1134

\bibitem[Bate et al. 2000]{} Bate, M. R., Bonnell, I. A., Clarke, C. J., Lubow, S. H., Ogilvie, G. I., Pringle, J. E., \& Tout, C. A. 2000, \mnras, 317, 773

\bibitem[]{} Bate, M. R., Lubow, S. H., Ogilvie, G. I., \& Miller, K. A. 2003, \mnras, 341, 213

\bibitem[]{} Beckwith, S. V. W., Sargent, A. I., Chini, R. S. \& Guesten, R. \aj, 1990, 99, 924

\bibitem[]{} Biretta, J., Kozhurina-Platais, V., Boffi, F., Sparks, W., \& Walsh, J. 2004, Instrument Science Report ACS 2004-09 (Baltimore: STScI) 

\bibitem[]{} Boss, A. P. 2002, \apj, 568, 743

\bibitem[Chen et al. 1990]{Chen90} Chen, W. P., Simon, M., Longmore, A. J., Howell, R. R., \& Benson, J. A. 1990, \apj, 357, 224

\bibitem[Close et al. 1998]{C98} Close, L. M., et al. 1998, \apj, 499, 883

\bibitem[]{} Connelley, M. S., Reipurth, B., \& Tokunaga, A. T. 2009, \aj, 138, 1193

\bibitem[]{} Ducati, J. R., Bevilacqua, C. M., Rembold, S. B., \& Ribeiro, D. 2001, \apj, 558, 309


\bibitem[]{} Dutrey, A., Guilloteau, S., Duvert, G., Prato, L., Simon, M., Schuster, K., \& Menard, F. 1996, \aap, 309, 493

\bibitem[]{} Eisl$\ddot{\mathrm{o}}$ffel, J. \& Mundt, R. 1998, \aj, 115, 1554

\bibitem[Elias 1978]{E78} Elias, J. H. 1978, \apj, 224, 857

\bibitem[]{} Ferland, G. \& Netzer, H. 1979, \apj, 229, 274

\bibitem[]{} Fukagawa, M., et al. 2004, \apj, 605, L53

\bibitem[Ghez et al. 1993]{G93} Ghez, A. M., Neugebauer. G., \& Mattews, K. 1993, \aj, 106, 5

\bibitem[]{} Gledhill, T. M. \& Scarrott, S. M. 1989, \mnras, 236, 139

\bibitem[G\"unther \& Kley 2002]{GK02} G\"unther, R. \& Kley, W. 2002, \aap, 387, 550

\bibitem[Hartigan \& Kenyon 2003]{HK03} Hartigan, P. \& Kenyon, S. J. 2003, \apj, 583, 334

\bibitem[Hawarden et al. 2001]{} Hawarden, T. G., Leggett, S. K., Letawsky, M. B., Ballantyne, D. R., \& Casali, M. M. 2001, \mnras, 325, 563	


\bibitem[Henyey \& Greenstein 1941]{HG41} Henyey, L. G. \& Greenstein, J. L., 1941, \apj, 93, 70

\bibitem[Hioki et al. 2007]{H07} Hioki, T., et al. 2007, \aj, 134, 880

\bibitem[Hioki et al. 2009]{H09} Hioki, T., et al. 2009, \pasj, 61, 1271

\bibitem[Hirth et al. 1997]{Hi97} Hirth, G. A., Mundt, R., \& Solf, J. 1997, \aaps, 126, 437

\bibitem[Itoh et al. 2002]{I02} Itoh, Y., et al. 2002, \pasj, 54, 963

\bibitem[Itoh et al. 2005]{I05} Itoh, Y., et al. 2005, \apj, 620, 984

\bibitem[]{} Jensen, E. L. N., Mathieu, R. D., Donar, A. X., \& Dullighan, A. 2004, \apj, 600, 789

\bibitem[]{} Kim, S.-H., Martin, P. G., \& Hendry, P. D. 1994, \apj, 422, 164

\bibitem[]{} Kitamura, Y., Momose, M., Yokogawa, S., Kawabe, R., Tamura, M., \& Ida, S. 2002, \apj, 581, 357

\bibitem[]{} Koekemoer, A. M. et al. 2007, \apjs, 172, 196

\bibitem[Krist et al. 1998]{K98} Krist, J. E., et al. 1998, \apjl, 501, 841

\bibitem[Krist et al. 2004]{K96} Krist, J. E. 2004, Tiny Tim User's Manual V 6.3
\bibitem[Krist et al. 2005]{K05} Krist, J. E., et al. 2005, \aj, 130, 2778

\bibitem[Krist et al. 2008]{K08} Krist, J. E., Stapelfeldt, K. R., Hester, J. J., Healy, K., Dwyer, S. J., \& Gardner, C. L. 2008, \aj, 136, 1980

\bibitem[]{} Kudo, T., et al. 2008, \apj, 673, L67

\bibitem[Leinert et al. 1993]{L93} Leinert, Ch., et al. 1993, \aap, 278, 129

\bibitem[]{} Lucas, P. W., et al. 2004, \mnras, 352, 1347

\bibitem[]{} Machida, M. N., Tomisaka, K., Matsumoto, T., \& Inutsuka, S. 2008, \apj, 677, 327

\bibitem[McCabe et al. 2002]{M02} McCabe,~C., Duch\^ene,~G., \& Ghez, A.M. 2002, \apj, 575, 974

\bibitem[Mathieu et al. 1997]{M97} Mathieu, R. D., et al., 1997, \aj, 113, 1841

\bibitem[]{} Mathis, J. S., Rumpl, W., \& Nordsieck, K. H. 1977, 217, 425

\bibitem[Mathis 1990]{Ma90} Mathis, J. S. 1990, aap, 28, 37

\bibitem[Murakawa et al. 2003]{M03} Murakawa, K., et al. 2003, \procspie, 4841, 881

\bibitem[]{} Nakamura, F. \& Li, Zhi-Yun. 2002, \apj, 566, 101

\bibitem[]{} Nelson, A. F., Benz, W., Adams, F. C., \& Arnett, D. 1998, \apj, 502, 342

\bibitem[Ochi et al. 2005]{O05} Ochi, Y., Sugimoto, K., \& Hanawa, T. 2005, \apj, 623, 922

\bibitem[]{} Pety, J., Gueth, F., Guilloteau, S., \& Dutrey, A. 2006, \aap, 458, 841

\bibitem[]{} Pinte, C., Fouchet, L., Menard, F., Gonzalez, J. F., \& Duch\^ene, G. 2007, \aap, 469, 963

\bibitem[]{} Potter, D. E., Close, L. M., Roddier, F., Roddier, C., Graves, J. E., \& Northcott, M. 2000, \apj, 540, 422

\bibitem[]{} Prato, L., Simon, M., Mazeh, T., Zucker, S., \&  McLean, I. S. 2002, \apj, 579, L99

\bibitem[]{} Roberge, A., Weinberger, A. J., \& Malumuth, E. M. 2005, \apj, 622, 1171

\bibitem[Schneider et al. 2003]{Sc03} Schneider, G., et al. 2003, \aj, 125, 1467

\bibitem[]{} Silber, J., Gledhill, T., Duch\^ene, G., \& Menard, F. 2000, \apj, 536, L89

\bibitem[]{} Simon, M., Chen, W. P., Howell, R. R., Benson, J. A., \& Slowik, D. 1992, \apj, 384, 212

\bibitem[]{} Simon, M., Close, L. M., \& Beck, T. L. 1999, \aj, 117, 1375

\bibitem[]{} Sirianni, M., et al. 2005, \pasp, 117, 1049

\bibitem[]{} Stapelfeldt, K. R., et al. 1998, \apj, 508, 736

\bibitem[]{} Stark, D. P., Whitney, B. A., Stassun, K., \& Wood, K. 2006, \apj, 649, 900

\bibitem[Takami et al. 2007]{} Takami, M., Beck, T. L., Pyo, T.-S., McGregor, P., \& Davis, Ch. 2007, \apj, 670, 33

\bibitem[Tamazian et al. 2002]{} Tamazian, V. S., Docobo, J. A., White, R. J., \& Woitas, J. 2002, \apj, 578, 925

\bibitem[]{} Tamura, M., \& Sato, S. 1989, \aj, 98, 4

\bibitem[Tamura et al. 2000]{T00} Tamura, M., et al. 2000, \procspie, 4008, 1153

\bibitem[]{} van de Hulst, H. C. 1957, Light Scattering by Small Particles (New York:Wiley)

\bibitem[]{} Vrba, F. J., Rydgren, A. E., \& Zak, D. S. 1985, \aj, 90, 2074

\bibitem[Weinberger et al. 2002]{W02} Weinberger, A., et al. 2002, \apj, 566, 409

\bibitem[]{} Weintraub, D. A., Goodman, A. A., \& Akeson, R. L. 2000, in Protostars and Planets IV, ed. V. Mannings, A. P. Boss, \& S. S. Russell (Tucson, AZ: Univ. Arizona Press), 247

\bibitem[]{} White, R. J., Ghez, A. M., Reid, I. N., \& Schultz, G. 1999, \apj, 520, 811

\bibitem[White \& Ghez 2001]{WG01} White, R. J. \& Ghez, A. M. 2001, \apj, 556, 265

\bibitem[]{} White, R. J., \& Hillenbrand, L. A. 2004, \apj, 616, 998

\bibitem[Whitney \& Hartmann 1992]{WH92} Whitney, B., \& Hartmann, L., 1992, \apj, 395, 529

\bibitem[]{} Whitney, B. A., Kenyon, S. J., \& Gomez, M. 1997, \apj, 485, 703

\bibitem[]{} Woitas, J., Eisl$\ddot{\mathrm{o}}$ffel, J., Mundt, R. \& Ray, T. P., 2002, \apj, 564, 834

\bibitem[]{} Wood, K., Crosas, M., \& Ghez, A. 1999, \apj, 516, 335

\end{thebibliography}
\end{document}